# AN EXTENSION OF THE STATISTICAL BOOTSTRAP MODEL TO INCLUDE STRANGENESS

A. S. Kapoyannis[*,+], C. N. Ktorides[*] and A. D. Panagiotou[*]

[*]*Division of Nuclear and Particle Physics, University of Athens, GR-15771 Athens, Hellas*

[+]*Institute of Nuclear Physics, NCSR Demokritos, GR-15310 Agia Paraskevi Attikis, Hellas*

## Abstract

We extend the Statistical Bootstrap Model (SBM) in order to describe hadronic systems which carry strangeness. We establish that the hadronic phase can exist only in a region of the phase space $(T,\mu_q,\mu_s)$, which is bounded by a "critical" surface. The SBM leads to a mass spectrum with the asymptotic form : $\rho(m) = m^{-\alpha} \exp[T/T^*]$, where the exponent $\alpha$ can take different values. We first employ the usual version of the SBM ($\alpha$=2) and impose strangeness neutrality in order to obtain the relation between the temperature and the chemical potentials $\mu_q$, $\mu_s$. We find an unphysical behaviour of $\mu_s$ near the critical surface. Subsequently, we study another version of SBM where $\alpha$=4 and we are led to a realistic as well as acceptable behaviour. In this case we are also able to relate the parameter of the model, $T_0$, (the critical temperature at zero $\mu_q$) with the MIT bag constant $B$. The relation between $(T,\mu_q,\mu_s)$ can be used to predict the particle ratios on the critical surface at the limit of the Hadron Gas phase.

## 1. Introduction

The Statistical Bootstrap Model (SBM) [1,2,3] constitutes an effort for a self consistent thermodynamical description of relativistic, multiparticle systems. The basic idea is to bypass the employment of interaction at a distance between particles, in favour of successive levels of organisation of matter into particle-like entities of increasing complexity known as **fireballs**. In the context of strong interaction physics the original set of **input** particles corresponds to all known hadrons.

The agent which carries dynamical information in the bootstrap scheme is the mass spectrum $\rho(m)$ of the fireballs. It satisfies an integral equation with the generic form [4]:

$$\rho(m) = \delta(m - m_0) + \sum_{n=2}^{\infty} \frac{1}{n!} \int \delta\left(m - \sum_{i=1}^{n} m_i\right) \prod_{i=1}^{n} \rho(m_i) dm_i \qquad (1)$$

where $m_0$ represents the mass of an input particle and $m_i$, i=1,2,...,∞, stands for the fireball masses in ascending order of complexity.

It was within the framework of the SBM that the idea of the existence of a critical temperature, beyond which the hadron phase of matter gives way to a different one, was first introduced [5]. Nowadays such a scenario receives concrete support from QCD, through lattice computations [6]. The emerging picture calls for a quark-gluon plasma (QGP) phase. Given the potential significance of strangeness as far as the identification of the QGP phase is concerned, we have undertaken the extension of the SBM so as to include the quantum number of strangeness.

## 2. Incorporation of Strangeness into the Bootstrap scheme

The generic form of the bootstrap equation in the presence of strangeness is the following:

$$\tilde{B}(p^2)\tilde{\tau}(p^2,b,s) = \underbrace{g_{bs}\tilde{B}(p^2)\delta_0(p^2 - m_{bs}^2)}_{\text{input term}} +$$

$$+ \sum_{n=2}^{\infty}\frac{1}{n!}\int \delta^4\left(p - \sum_{i=1}^{n}p_i\right)\sum_{\{b_i\}}\delta_K\left(b - \sum_{i=1}^{n}b_i\right)\sum_{\{s_i\}}\delta_K\left(s - \sum_{i=1}^{n}s_i\right) \cdot$$

$$\cdot \prod_{i=1}^{n}\tilde{B}(p_i^2)\tilde{\tau}(p_i^2,b_i,s_i)d^4p_i \quad , \tag{2}$$

where $b(s)$ stands for baryon (strangeness) number and the $g_{bs}$ represent the degeneracy factors.

Let us provide the main ingredients which have been employed in order to reach the above result. We use a system of units in which $\hbar = c = k = 1$ and introduce Touscheck's integration measure [7], according to which

$$\frac{V}{h^3}d^3p \cdot \int \rho(m,b,s)dm \to \frac{2V^\mu p_\mu}{(2\pi)^3}\int \delta_0(p^2 - m^2)d^4p \cdot \tau(m^2,b,s)dm^2, \tag{3}$$

where $V^\mu$ is a boosted four-volume, which is parallel to the four-momentum of a given particle:

$$V^\mu = \frac{V}{m}p^\mu \tag{4}$$

This relation underlies the specification that each fireball carries its own volume [8].Note that the expression on the left side of the arrow in (3) counts the number of available particle states in the rest frame of volume V, which are characterised by the mass spectrum $\rho(m)$.

We have further converted $\rho(m)$ to the spectrum function $\tau(m^2)$ according to

$$\rho(m)dm = \tau(m^2)dm^2$$

and have set

$$B(p^2) = \frac{2V^\mu p_\mu}{(2\pi)^3} = \frac{2Vm}{(2\pi)^3} \ . \tag{5}$$

We have also introduced a rearrangement of the form

$$B(p^2)\tau(p^2,b,s) \equiv \tilde{B}(p^2)\tilde{\tau}(p^2,b,s) \ . \tag{6}$$

The specific choice one makes in using the above relation is of crucial significance as far as the dynamical description of the SBM is concerned. Next we carry out three Laplace transformations, one continuous and two discrete, which lead to the replacements:

$$(P_\mu, b, s) \to (\beta_\mu, \lambda_B, \lambda_S) \ .$$

The new set of variables will be vested with a thermodynamical content in terms of the inverse (four) temperature and the fugacities for baryon and strangeness quantum number, respectively.

In the centre of mass frame and with the transformation $\lambda_q = \lambda_B^{1/3}$, $\lambda_s = \lambda_B^{1/3}\lambda_S^{-1}$, which introduces the quark fugacities $\lambda$ ($q \equiv$ up-down, $s \equiv$ strange), we arrive at the following form for the bootstrap equation

$$\varphi(\beta, \lambda_q, \lambda_s) = 2G(\beta, \lambda_q, \lambda_s) - \exp[G(\beta, \lambda_q, \lambda_s)] + 1 \ , \tag{7}$$

where $\varphi(\beta, \lambda_q, \lambda_s)$ is the known **input term**:

$$\varphi(\beta, \lambda_B, \lambda_S) = \sum_{b=-\infty}^{\infty} \lambda_B^b \sum_{s=-\infty}^{\infty} \lambda_S^s \int e^{-\beta^\mu p_\mu} g_{bs} \tilde{B}(p^2) \delta_0(p^2 - m_{bs}^2) d^4p$$

$$= \frac{2\pi}{\beta} \sum_{b=-\infty}^{\infty} \lambda_B^b \sum_{s=-\infty}^{\infty} \lambda_S^s g_{bs} \tilde{B}(m_{bs}^2) m_{bs} K_1(\beta m_{bs}) \tag{8}$$

and $G(\beta, \lambda_q, \lambda_s)$ is the **mass-spectrum containing term**:

$$G(\beta, \lambda_B, \lambda_S) = \sum_{b=-\infty}^{\infty} \lambda_B^b \sum_{s=-\infty}^{\infty} \lambda_S^s \int e^{-\beta^\mu p_\mu} \tilde{B}(p^2) \tilde{\tau}(p^2, b, s) d^4 p$$

$$= \frac{2\pi}{\beta} \int_0^\infty m\tilde{B}(m^2) \tilde{\tau}(m^2, \lambda_B, \lambda_S) K_1(\beta m) dm^2 , \qquad (9)$$

which involves the unknown function $\tilde{\tau}$. In eqs. (8) and (9) $K$ denotes the modified Bessel function of the second kind.

The bootstrap equation (7) displays, in the $\varphi - G$ plane, Fig. 1, a square root branch point at

$$\varphi(T_{cr}, \mu_{q\,cr}, \mu_{s\,cr}) = \ln 4 - 1 , \qquad (10a)$$

$$G(T_{cr}, \mu_{q\,cr}, \mu_{s\,cr}) = \ln 2 . \qquad (10b)$$

Eq. (10a) defines a **critical surface** in the 3-d space $(T, \mu_q, \mu_s)$ which sets the limits of the hadronic phase. Points radially outside the critical surface belong to unphysical solutions of the bootstrap equation and are thereby assigned to a region where a new phase of matter, presumably the QGP phase, makes its appearance.

Let us also introduce the temperature $T_0$ according to

$$\varphi(T_0, \lambda_q = 1, \lambda_s = 1) = \ln 4 - 1 , \qquad (11)$$

which constitutes the highest temperature beyond which the Hadron Gas phase does not exist.

We finally turn our attention to the thermodynamical description of the system. To this end we take into account the fact that according to the bootstrap scheme the number of available states in a volume $d^3 p$ around $\vec{p}$, baryon number $B$ and strangeness number $S$ is given, in covariant form, by

$$\frac{2V_\mu^{ext} p^\mu}{(2\pi)^3} \tilde{\tau}(p^2,b,s) d^4p \ , \tag{12}$$

where $V_\mu^{ext}$ is the total external (four) volume available to the system. It is a constant as far as the integration over $d^4p$ is concerned. Accordingly, the grand canonical partition function for the system reads, in covariant form,

$$\ln Z(\beta, V, \lambda_B, \lambda_S) = \sum_{b=-\infty}^{\infty} \lambda_B^b \sum_{s=-\infty}^{\infty} \lambda_S^s \int \frac{2V_\mu p^\mu}{(2\pi)^3} \tilde{\tau}(p^2,b,s) e^{-\beta^\mu p_\mu} d^4p \ .$$

Switching to quark fugacities, choosing the four-vectors $V^\mu$ and $\beta^\mu$ to be parallel and going to the frame for which $\beta^\mu = (\beta,0,0,0)$ [8] we write

$$\ln Z(\beta, V, \lambda_q, \lambda_s) = \frac{V}{\beta 2\pi^2} \cdot \int_0^\infty m^2 \tilde{\tau}(m^2, \lambda_q, \lambda_s) K_2(\beta m) dm^2 \ . \tag{13}$$

Our problem now is to express the above partition function in terms of the function $G(\beta, \lambda_q, \lambda_s)$ which contains the bootstrap mass spectrum. Once this is done we shall be in position to extract specific results from the extended SBM, via the inclusion of strangeness.

## 3. Connection with Phenomenology and Results

Let us return to equation (4). The (fireball) volume to mass constant provides a quantity that can be related to the MIT bag model [9]. We set

$$\frac{V}{m} = \frac{V_i}{m_i} = \frac{1}{4B} \ , \tag{14}$$

where $B$ is the MIT bag constant and where $V_i(m_i)$ denotes the volume (mass) of the fireball. The first equality in (14) comes from the assumption that the volume (mass) of a given fireball is the sum of the volumes (masses) of the constituent fireballs.

Let us assess the splitting between the $B(p^2)$ and $\tau(p^2,b,s)$ in the bootstrap equation. We start with the "natural" definition of $B(p^2)$ as given by (6). Here, we have a purely kinematical assignment to this quantity so all dynamics of the bootstrap model are carried by $\tau(p^2,b,s)$ [10]. Setting $B(m^2) \equiv H_0 m^2$ we find in this case

$$H_0 = \frac{2}{(2\pi)^3 4B} \ . \qquad (15)$$

A rearrangement of the factors $\tilde{B}$ and $\tilde{\tau}$ would imply a behaviour of the form $\tilde{B}(m^2) = const \cdot m^d$. Any choice for which $d \neq 2$ entails an absorption of part of the dynamics into $\tilde{B}$. Traditionally, SBM applications have centered around the choice $\tilde{B}(m^2) \sim m^4$. Setting $B(m^2) \equiv H_2 m^4$ we are now obliged to introduce a reference mass scale $\tilde{m}$ in order to relate $\tilde{\tau}$ with $\tau$:

$$\tilde{\tau}(m^2,b,s) = \frac{m^2}{\tilde{m}^2} \tau(m^2,b,s) \ . \qquad (16)$$

We also determine, for this case,

$$H_2 = \frac{2\tilde{m}^2}{(2\pi)^3 4B} \ . \qquad (17)$$

We stress that for any choice other than the one given by (6), one is forced to enter a reference mass scale into SBM descriptions.

Given the above remarks, relevant to the phenomenological connection with QCD, let us turn our attention to the asymptotic behaviour of the mass spectrum function $\rho(m)$, as $m \to \infty$. It can be shown [11,12] that

$$\tilde{B}(m^2)\tilde{\tau}(m^2,\{\lambda\}) \xrightarrow[m \to \infty]{} C(\{\lambda\}) m^{-3} \exp[m/T^*(\{\lambda\})] \ , \qquad (18)$$

where $T^*(\{\lambda\})$ satisfies the criticality equation, c.f. Eq. (10a). In the above relation $\{\lambda\}$ is a collective index for fugacities, while $C(\{\lambda\})$ is a quantity independent of mass.

For a given choice $\tilde{B}(m^2) = const \cdot m^d$ we have

$$\tilde{\tau}(m^2,\{\lambda\}) \xrightarrow[m\to\infty]{} C'(\{\lambda\}) m^{-c} \exp[m/T^*(\{\lambda\})] , \qquad (19)$$

where c=3+d.

In the literature it is more common to refer to the index $\alpha = c - 1$, which gives the exponent entering the asymptotic behaviour of $\tilde{\rho}(m,\{\lambda\})$. Thus the choices, entailed by relations (15) and (17), correspond to $\alpha = 4$ and $\alpha = 2$, respectively. These, in fact, happen to be the two cases which facilitate analytic procedures linking the canonical partition function to the term $G$, which contains the mass spectrum, and eventually, through the bootstrap equation, to the input term $\varphi$. We find, for $\alpha = 2$ [13,14],

$$\ln Z(V,\beta,\lambda_q,\lambda_s) = -\frac{2V}{(2\pi)^3 H_2} \frac{\partial}{\partial \beta} G(\beta,\lambda_q,\lambda_s)$$

$$= -\frac{V}{4\pi^3 H_2} \cdot \frac{1}{2 - e^{G(\beta,\lambda_q,\lambda_s)}} \frac{\partial \varphi(\beta,\lambda_q,\lambda_s)}{\partial \beta} \qquad (20)$$

and for $\alpha = 4$ [14],

$$\ln Z(\beta,V,\lambda_q,\lambda_s) = \frac{V}{4\pi^3 H_0} \frac{1}{\beta^3} \int_\beta^\infty x^3 G(x,\lambda_q,\lambda_s) dx . \qquad (21)$$

Now, hadronic interactions pertain to physical situation where the total strangeness number is zero. In the thermodynamic context of the SBM the relevant condition is

$$\langle S \rangle = \lambda_s \frac{\partial \ln Z(V, \beta, \lambda_q, \lambda_s)}{\partial \lambda_s}\bigg|_{(V, \beta, \lambda_q)} = 0 . \tag{22}$$

The above relation determines a surface in the $(T, \mu_q, \mu_s)$ space. For $\alpha = 2$, we determine

$$\frac{e^{G(\beta, \lambda_q, \lambda_s)}}{\left[2 - e^{G(\beta, \lambda_q, \lambda_s)}\right]^2} \cdot \frac{\partial \varphi(\beta, \lambda_q, \lambda_s)}{\partial \lambda_s} \cdot \frac{\partial \varphi(\beta, \lambda_q, \lambda_s)}{\partial \beta} + \frac{\partial^2 \varphi(\beta, \lambda_q, \lambda_s)}{\partial \lambda_s \partial \beta} = 0 \tag{23}$$

and for $\alpha = 4$,

$$\int_0^T \frac{1}{y^5} \frac{1}{2 - \exp[G(y, \lambda_q, \lambda_s)]} \frac{\partial \varphi(y, \lambda_q, \lambda_s)}{\partial \lambda_s} dy = 0 . \tag{24}$$

The derivations of the above equations along with numerical approaches to their solutions will be given elsewhere [14].

Before we present results from our studies of the critical and the $\langle S \rangle = 0$ surfaces for the cases $\alpha = 2$ and $\alpha = 4$ we should comment on the following potential disparity between them. According to the analysis conducted in [8] these two values for $\alpha$ lie below and above the threshold value of $\alpha = 7/2$, respectively. This separates the regime where the energy density of the hadronic system goes to infinity, as $T \to T_{cr} (\alpha < 7/2)$ and where it stays finite $(\alpha \geq 7/2)$. On this basis, the $\alpha = 4$ value appears to be the most desirable one, as far as the existence and observability of a QGP phase is concerned. On the other hand Hagedorn and Rafelski [13] have shown that the infinity problem for $\alpha < 7/2$ can be bypassed through finite volume corrections. In fact, it was on this basis that the $\alpha = 2$ choice has dominated the theoritical scene as far as the SBM is concerned.

In Fig. 2a we depict the profile of the critical surface for $\alpha = 2$ and $\alpha = 4$ by displaying two characteristic cuts of this surface perpendicular to the $\mu_s$ axis. Fig. 2b depicts a similar setup with the cuts now occuring perpendicularly to the $\mu_q$ axis. It is observed that the results depicted in the above figures seem to advocate similar qualitative behaviour between the two choices of $\alpha = 2$ and $\alpha = 4$.

A radical departure between the two values of $\alpha$ appears as soon as one compares the way by which the $\langle S \rangle = 0$ surface meets the critical one. Leaving the discussion of the numerical methodology which produces the corresponding curves to our forthcoming paper [14], let us explain the way we have chosen to display the relevant results. In Fig. 3 we show how, for different values of $T_0$, the $\langle S \rangle = 0$ surface joins the critical surface along the direction $\mu_q/T = 0.4$ in the $\mu_s - T$ plane. The almost vertical lines represent the profile of the critical surface near the $\mu_q - T$ plane (the $\mu_q$ axis is not shown) which is, in fact, cylindrical. For $\alpha = 2$ (Fig. 3a) we have the unphysical behaviour of a rising chemical potencial $\mu_s$ as we approach the end of the hadronic phase. By juxtaposition the meshing between the two surfaces for $\alpha = 4$ (Fig. 3b) has a desirable, as well as reasonable behaviour. It is on this basis that we consider the $\alpha = 4$ version as the most realistic one, as far as the approach to the QGP phase from the hadronic side is concerned and it is for this case that we intend to focus our SBM-based analysis on particle ratios.

For comparison we have added to the same diagram the intersection of the plane $\mu_q/T = 0.4$ with the surface $\langle S \rangle = 0$, which comes from solving the equation

$$F_K(T)\left(\lambda_s \lambda_q^{-1} - \lambda_s^{-1} \lambda_q\right) + F_{Hyp}(T)\left(\lambda_s \lambda_q^{2} - \lambda_s^{-1} \lambda_q^{-2}\right) + \\ 2F_\Xi(T)\left(\lambda_s^{2} \lambda_q - \lambda_s^{-2} \lambda_q^{-1}\right) + 3F_\Omega(T)\left(\lambda_s^{3} - \lambda_s^{-3}\right) = 0 \quad . \tag{25}$$

The above equation is derived for an ideal hadron gas (see for example [15,16]).

Fig. 4a(b) depicts the projection of the intersection of the surface $\langle S \rangle = 0$ with the critical surface for different values of $T_0$ on the $\mu_s - \mu_q$ plane, when $\alpha = 2$ ($\alpha = 4$).

Fig. 5 gives a 3-d picture of how $\langle S \rangle = 0$ curves along different directions in the $\mu_q - T$ plane and intersects with the critical surface for a specific value of $T_0$, when $\alpha = 4$.

Finally in the framework $\alpha = 4$ model one can directly relate $T_0$ and the bag model parameter $B$. In table 1 we show how the corresponding values pair up. On a phenomenological basis one has the option of using as input either a value for $T_0$ or a value for $B$, an occurrence which further underlines the advantages of the $\alpha = 4$ model.

| $T_0$ (MeV) | $B^{1/4}$ (MeV) |
|---|---|
| 150 | 154.690 |
| 160 | 177.120 |
| 170 | 201.170 |
| 180 | 226.704 |
| 190 | 253.578 |
| 200 | 281.655 |

Table 1. Connection between $T_0$ and B.

## 4. Summary and Conclusions.

In this paper, we presented an enlargement of the SBM scheme which includes the strangeness quantum number. Our efforts were directed towards the determination, in the 3-dimensional $(T, \mu_q, \mu_s)$ space, of : 1) the critical surface which sets the limits of the hadronic phase and which is determined by the bootstrap equation and 2) the $\langle S \rangle = 0$ surface, relevant to hadronic processes, which is determined from the partition function. Finally we compared two specific SBM versions corresponding to the values $\alpha = 2$ and $\alpha = 4$ of the exponent, which enters the asymptotic expression for the mass spectrum $\rho(m)$. We found desicive advantages of the second ($\alpha = 4$), as opposed to the first version which has dominated SBM studies in the past. Note that the $\alpha = 4$ version admits, as direct phenomenological input from QCD, the value of the MIT bag constant and relates it uniquely to $T_0$ .

*Acknowledgement.* We wish to thank Prof. R. Hagedorn for useful and interesting discussions.

## Figures

Fig. 1. $G$ as a function of $\varphi$.

Fig. 2. (a) Comparison of the intersections of planes of constant s-quark chemical potential $\mu_s$ with the critical surface $\varphi(T, \mu_q, \mu_s) = \ln 4 - 1$ for the values of $\alpha = 2$ and $\alpha = 4$ and for $T_0$=180 MeV.  (b) Comparison of the intersections of planes of constant q-quark chemical potential $\mu_q$ with the critical surface $\varphi(T, \mu_q, \mu_s) = \ln 4 - 1$ for the values of $\alpha = 2$ and $\alpha = 4$ and for $T_0$=180 MeV.

Fig. 3. (a) Projection on the plane $(T, \mu_s)$ of intersections of planes of constant $\lambda_q$ ($\mu_q/T = 0.4$) with the surface $\langle S \rangle = 0$ for different values of $T_0$, when $\alpha = 2$.  (b) Projection on the plane $(T, \mu_s)$ of intersections of planes of constant $\lambda_q$ ($\mu_q/T = 0.4$) with the surface $\langle S \rangle = 0$ for different values of $T_0$, when $\alpha = 4$.

Fig. 4. (a) Projection on the plane $(\mu_q, \mu_s)$ of the intersection of the critical surface and the surface $\langle S \rangle = 0$, when $\alpha = 2$.  (b) Projection on the plane $(\mu_q, \mu_s)$ of the intersection of the critical surface and the surface $\langle S \rangle = 0$, when $\alpha = 4$.

Fig. 5. Intersections of planes of constant $\lambda_q$ with the surface $\langle S \rangle = 0$ and the intersection of this surface with the critical one, for $T_0$=185.655 MeV, when $\alpha = 4$.

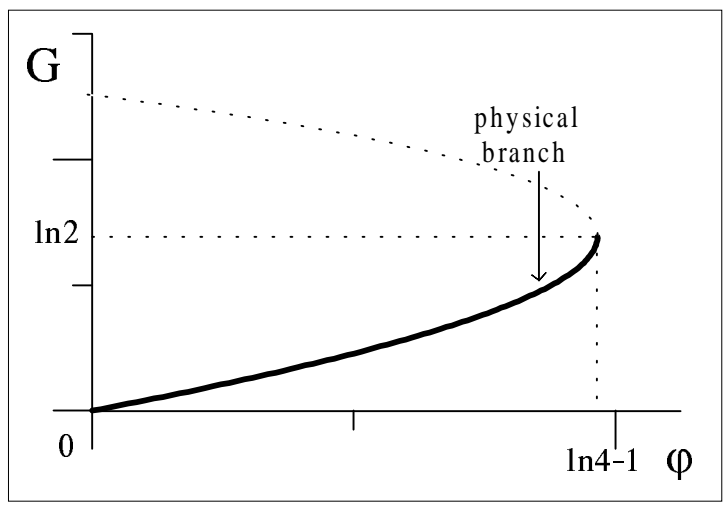

Fig.1

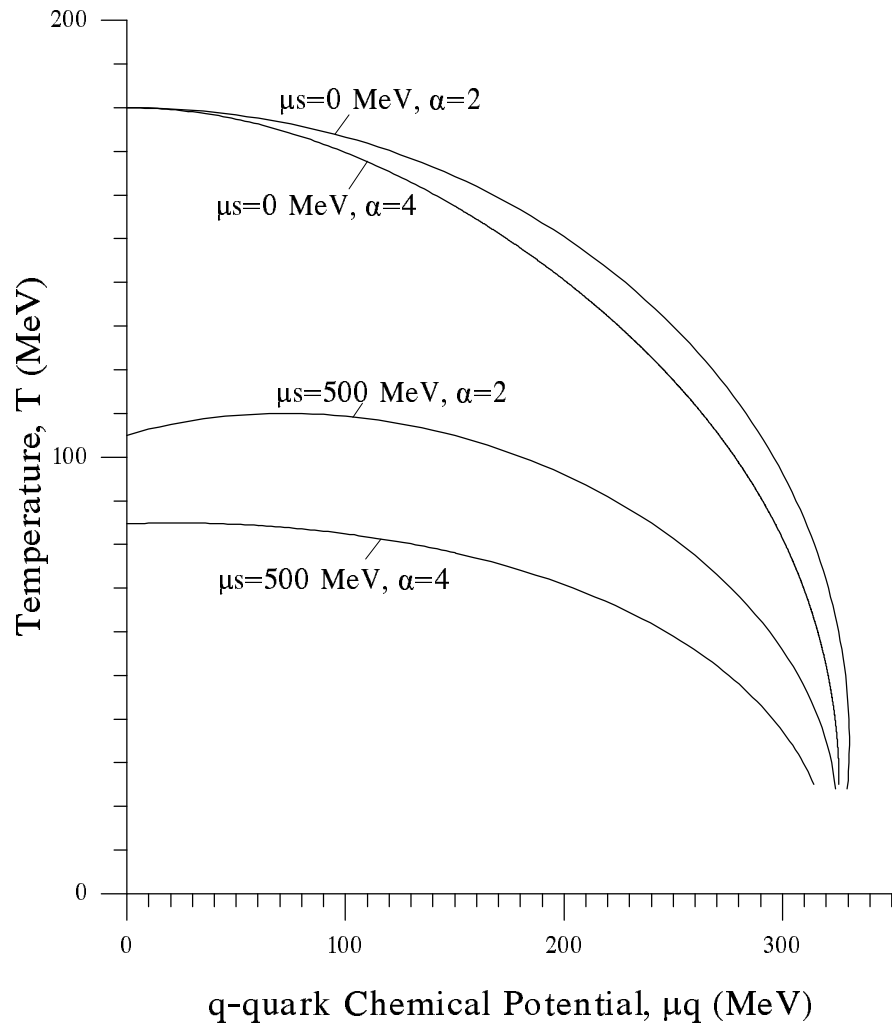
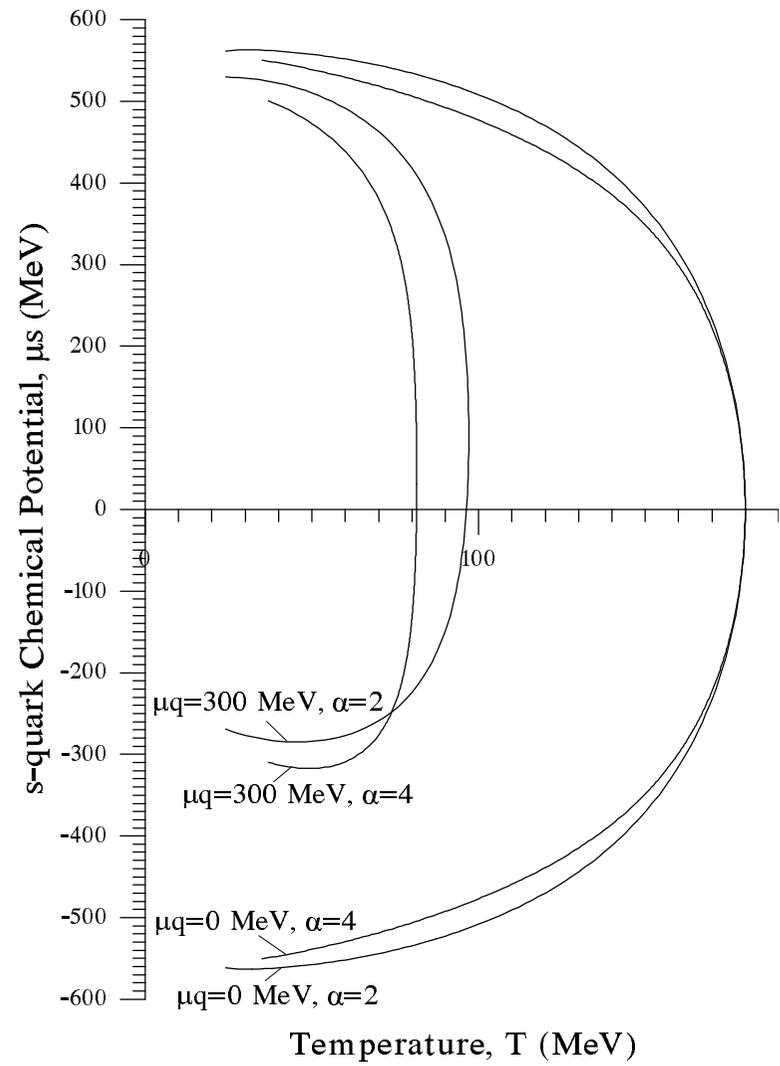

Fig. 2a

Fig. 2b

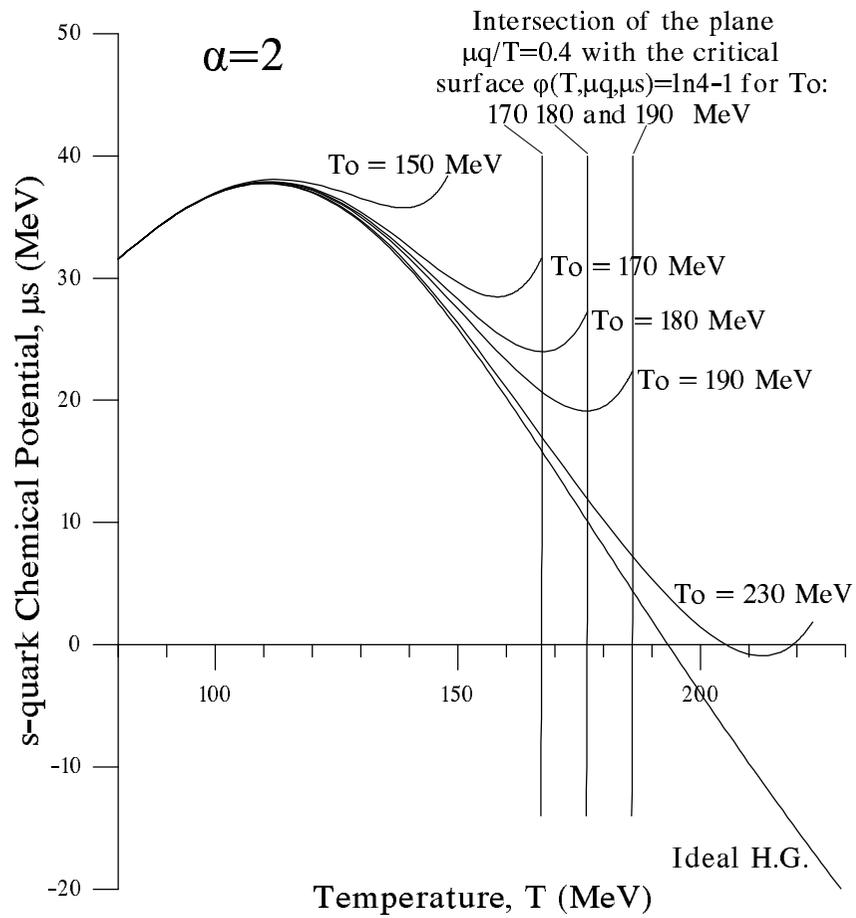

Fig. 3a

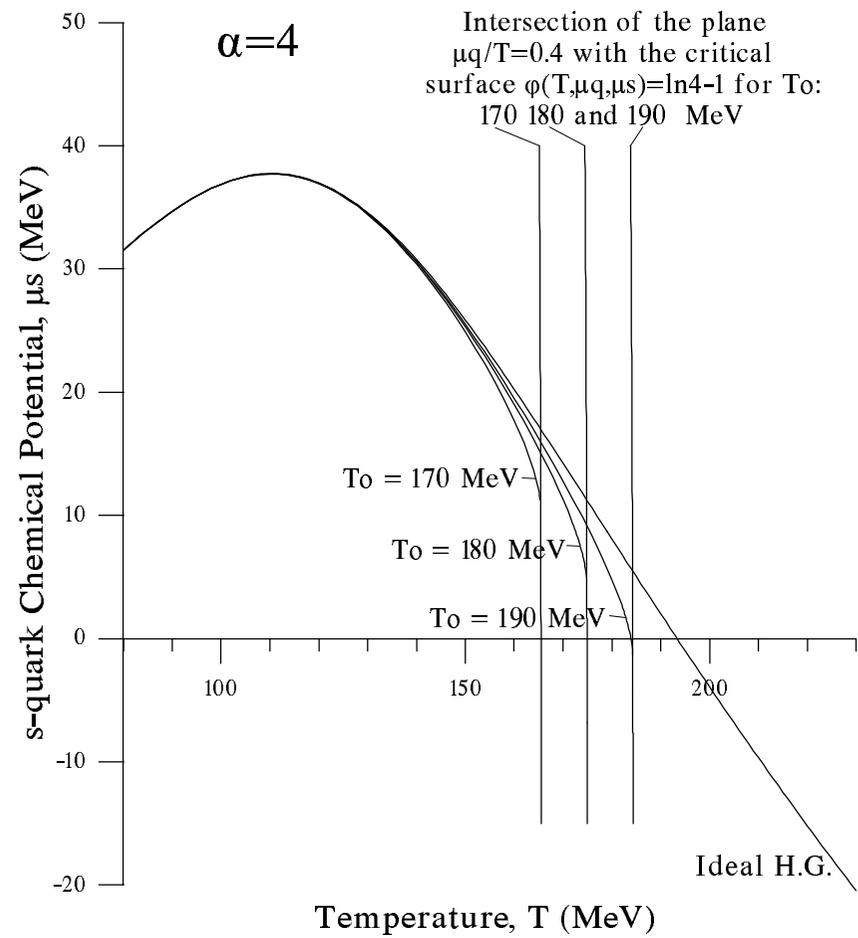

Fig. 3b

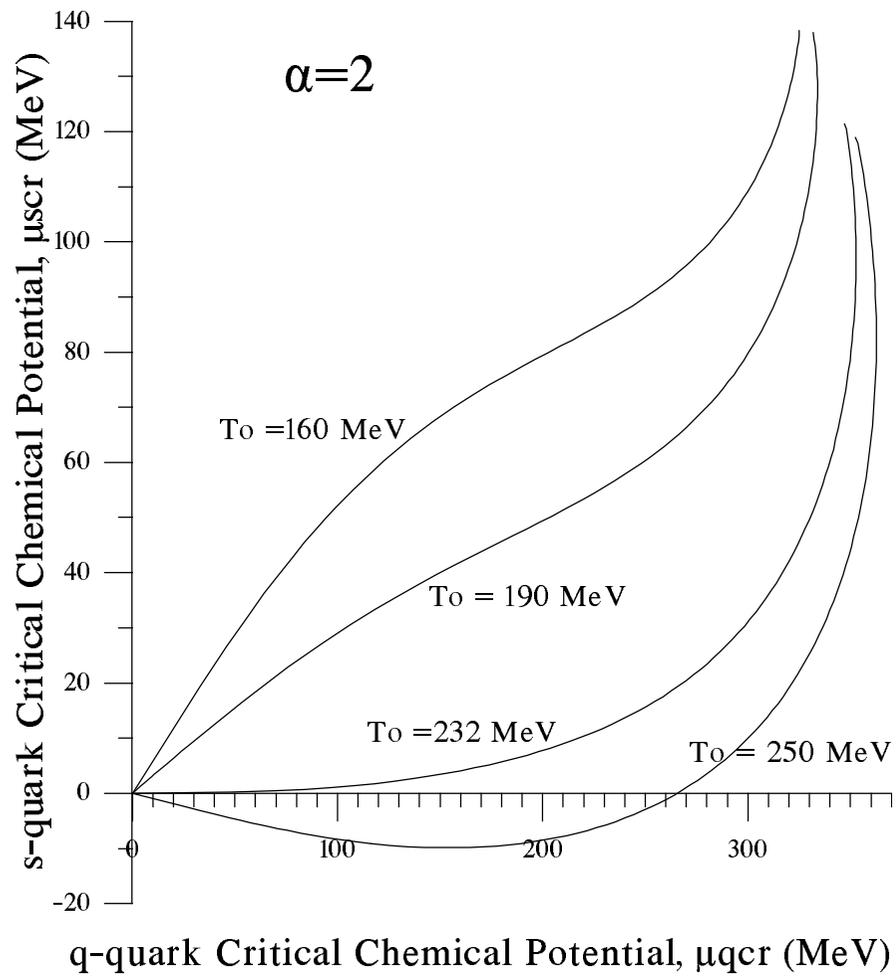 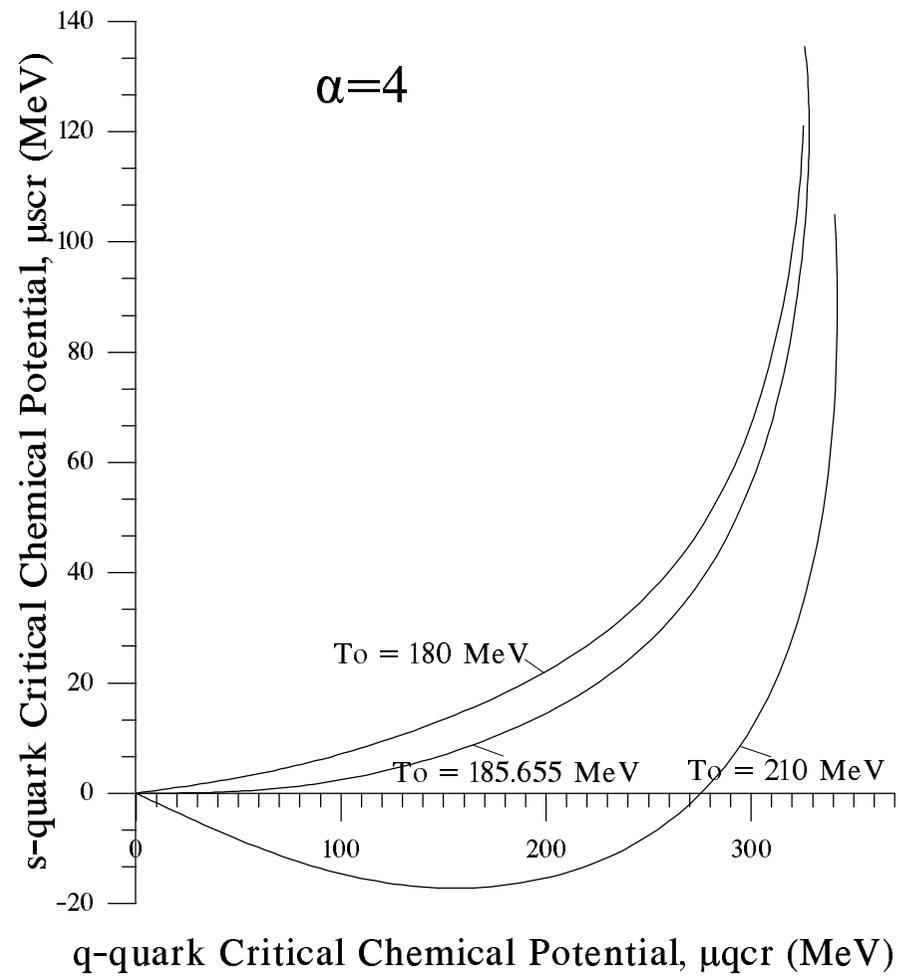

Fig. 4a

Fig. 4b

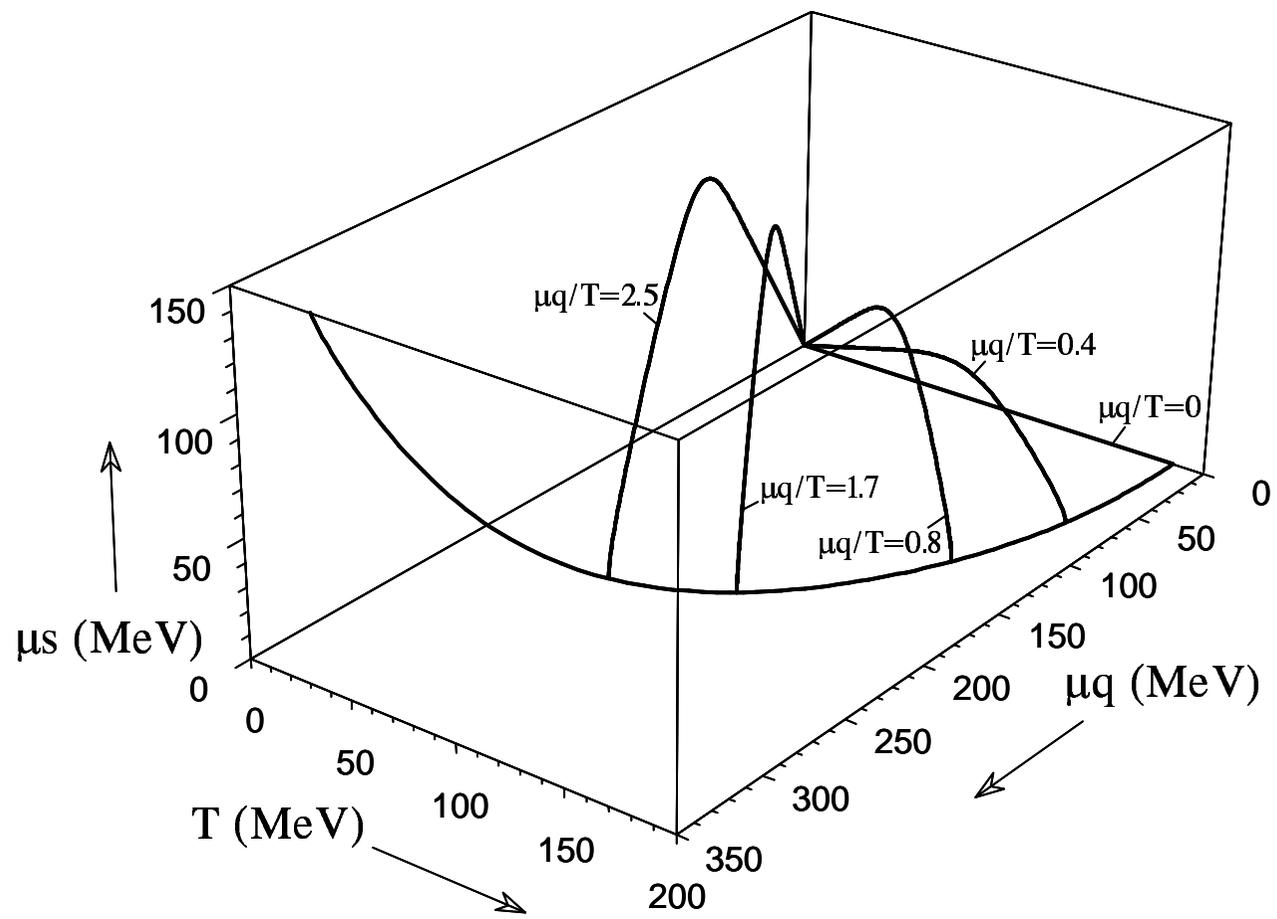

Fig. 5